\documentclass[%
aip,
amsmath,amssymb,
reprint,%
]{revtex4-1}

\usepackage{blindtext}
\usepackage{dcolumn}
\usepackage{bm}
\usepackage{color, soul, colortbl} 
\usepackage[colorlinks=true,
linkcolor=blue,
urlcolor=blue,
citecolor=blue,
bookmarks=true,
pdfborder={0 0 0}]{hyperref}

\usepackage[utf8]{inputenc}
\usepackage[T1]{fontenc}

\usepackage[rgb]{xcolor}

\usepackage{mdframed} 

\usepackage{graphicx,epsfig}
\usepackage{subfigure}
\usepackage{amsmath}

\usepackage{mathtools}
\usepackage{parskip}

\draft 

\begin{document}

\title{Multilayer modeling of adoption dynamics in energy demand management}

\author{Iacopo Iacopini}
\email[E-mail: ]{i.iacopini@qmul.ac.uk}
\affiliation{School of Mathematical Sciences, Queen Mary University of London, London E1 4NS, United Kingdom}
\affiliation{Centre for Advanced Spatial Analysis, University College London, London, W1T 4TJ, United Kingdom}
\affiliation{The Alan Turing Institute, The British Library, London NW1 2DB, United Kingdom}
\author{Benjamin Sch{\"a}fer}
\affiliation{School of Mathematical Sciences, Queen Mary University of London, London E1 4NS, United Kingdom}
\affiliation{
Chair for Network Dynamics, Center for Advancing Electronics Dresden (cfaed) and Institute for Theoretical Physics, TU Dresden, 01062 Dresden, Germany}
\author{Elsa Arcaute}
\affiliation{Centre for Advanced Spatial Analysis, University College London, London, W1T 4TJ, United Kingdom}
\author{Christian Beck}
\affiliation{School of Mathematical Sciences, Queen Mary University of London, London E1 4NS, United Kingdom}
\author{Vito Latora}
\email[E-mail: ]{v.latora@qmul.ac.uk}
\affiliation{School of Mathematical Sciences, Queen Mary University of London, London E1 4NS, United Kingdom}
\affiliation{The Alan Turing Institute, The British Library, London NW1 2DB, United Kingdom}
\affiliation{Dipartimento di Fisica ed Astronomia, Università di Catania and INFN, I-95123 Catania, Italy}
\affiliation{Complexity Science Hub Vienna (CSHV), 1080 Vienna, Austria}

\begin{abstract}
Due to the emerging of new technologies, the whole electricity system
is undergoing transformations on a scale and pace never observed
before. The decentralisation of energy resources and the smart grid
have forced utility services to rethink their relationships with
customers. Demand response (DR) seeks to adjust the demand for power
instead of adjusting the supply. However, DR business models rely on
customer participation and can only be effective when large numbers of
customers in close geographic vicinity, e.g. connected to the same
transformer, opt in.
Here, we introduce a model for the dynamics of service adoption on 
a two-layer multiplex network: the layer of social interactions among customers and the power-grid layer
connecting the households. While the adoption process\textemdash based on peer-to-peer communication\textemdash runs on the social layer, the time-dependent recovery rate of the nodes depends on the states of
their neighbors on the power-grid layer, making
an infected node surrounded by infectious ones less keen to recover. 
Numerical simulations of the model on synthetic and real-world
networks show that a strong local influence of the customers' actions
leads to a discontinuous transition where either none or all the nodes
in the network are infected, depending on the infection rate and the
social pressure to adopt. We find that clusters of early adopters act
as points of high local pressure, helping maintaining adopters, and
facilitating the eventual adoption of all nodes. This suggests direct
marketing strategies on how to efficiently establish and maintain new
technologies such as DR schemes.
\end{abstract}

\maketitle 

\begin{quotation} 
The electricity system is in the midst of large transformations, and new business models have emerged quickly to facilitate new modes of operation of the electricity supply. The so-called demand response seeks to coordinate demand from a large number of users, through incentives which are usually economic such as variable pricing tariffs. 
Here, we propose a simple mathematical framework to model consumer behaviors under
demand response. Our model considers at the same time social influence and customer benefits 
to opt into and stay within new control schemes. 
In our model information about the existence of a contract 
propagates through the links of a social network, while the geographic proximity of clusters of adopters influences the likelihood of participation by decreasing the likelihood of opting out.
The results of our work can help to take informed decision in energy demand management.
\end{quotation}

\section{Introduction}
The study of dynamical processes on complex networks is a well established branch of complex systems science that aims at understanding the complex interplay between the dynamics of the process and the topology of the underlying network \cite{porter2006dynamical, barrat2008dynamical}.
Networks encompass a powerful approach, in which a system can be represented by considering its connectivity patterns, encoding in this way all the interactions between the different units composing it into a compact framework \cite{albert2002statistical, newman2003structure, latora_nicosia_russo_2017}. Systems composed by units that interact in different ways can be analogously represented by considering their multilayered interactions \cite{de2013mathematical, boccaletti2014structure, kivela2014multilayer, battiston2017new,bianconi2018multilayer}. The highly versatile essence of the network representation allows one to use it as a structure for processes of very different nature, that can ultimately be used to model real world phenomena.
Among the most studied processes on networks, together with synchronization \cite{arenas2008synchronization} and random walks \cite{masuda2017random}, are the dynamics of spreading phenomena in a population, such as the spreading of diseases \cite{pastor2015epidemic}, norms, innovation adoption \cite{valente1996network,iacopini2018network} or knowledge diffusion \cite{cowan2004network}. Adoption dynamics becomes increasingly relevant when implementing new business models e.g. for the Internet of Things or smart grids \cite{Fang2012}.

When dealing with adoption or spreading processes, the typical approach 
is to divide the individuals of a population into a finite number 
of classes, or compartments. In the simplest possible case we have 
only two classes. Individuals or agents can either be in the susceptible (S) or in the infected (I) class, 
with the latter being the class of those who have an infection or have adopted a technology and are therefore potentially contagious for the rest of the population. Here, we focus on the SIS model, one 
of the simplest compartmental models that can be built as it uses just these two classes. The SIS is indeed suitable for modelling the dynamics of those diseases that can infect an individual more than once, such as common cold. 
In the SIS model all individuals are initially assigned to the S class, with the exception of a small initial seed of infected nodes. Infected individuals can then pass the infection to the susceptible ones by means of contacts, i.e., through the links of the network. More precisely, an infectious node can pass the infection to a neighbor according to a given rate of infection $\beta$. In turn, infected individuals spontaneously recover with a rate $\mu$ and then can get infected again.
This contagion dynamics goes typically under the name of {\it simple contagion}, to  stress the fact that a susceptible node in contact with more than one infected neighbor can get an infection by means of independent exposures. 
The modelling approach just described is not only restricted to the spreading of viruses, it can also cover a broader class of phenomena, such as smart-grid technologies, or the spread of behaviors like obesity \cite{christakis2007spread}, happiness \cite{fowler2008dynamic}, or smoking \cite{christakis2008collective}.

However, when dealing with phenomena that involve social contagion, it turns out that sometimes the simple contagion framework is not 
the most appropriate way to model the system under study. This is because the standard SIS model does not capture the basic dynamics of social influence and reinforcement, nor the non-linear nature of technological learning/adoption processes \cite{pinheiro2014origin, hodas2014simple}. This has been confirmed by relatively recent investigations on adoption patterns in online social networks \cite{centola2010spread,ugander2012structural,karsai2014complex,monsted2017evidence,iniguez2018service}. Therefore, {\it complex contagion} \cite{centola2007complex, guilbeault2018complex} has been proposed as an alternative description in which, for example,  threshold mechanisms are introduced in order to account for the effects of peer pressure and social reinforcement mechanisms \cite{bottcher2017critical}. The fundamental difference between simple and complex contagions relies on the fact that in the latter setting multiple exposures from different sources are required for a transmission event to happen. This idea has also been further extended in the recently introduced {\it simplicial contagion} model, in which a simplicial complex instead of a  graph is used as the underlying structure of a social systems to encode high-order (higher than pairwise) interactions among individuals 
\cite{iacopini2019simplicial}.
Another way of including social effects into the contagion process consists either in allowing the dynamics of infection to depend on some local properties of the node and their neighborhood, or alternatively in  letting nodes control for their connections \cite{gross2006epidemic, zanette2008infection, risau2009contact, gomez2016explosive, tuzon2018continuous, lee2019social}. Ultimately, the introduction of local effects into the contagion dynamics allows to effectively introduce mechanisms of awareness \cite{funk2009spread, funk2010endemic, wu2012impact, granell2013dynamical,steinegger2018interplay}, trust \cite{wu2017influence},  and risk perception \cite{bagnoli2007risk}.

All the models mentioned above focus on one of the two aspects of the dynamics of a spreading process, that is the contagion mechanism. This is generally controlled by means of the infection parameter $\beta$, which might eventually be node-dependent if local effects are considered. In the case of simple contagion the parameter $\beta$
mediates two-body interactions, with a corresponding process $S+I \to 2I$, while in the case of complex and simplicial contagion one-to-many-body and group interactions are considered respectively. Conversely, less attention has been devoted to the other aspect, that is the recovery mechanism.
The recovery rate parameter $\mu$ is typically considered constant for all the nodes, and it is usually absorbed into an effective infection rate $\beta/\mu$. Nevertheless, recent results have shown that heterogeneity in recovery rates can have dramatic effects on the type and position of epidemic transitions, implying that heterogeneous infectious periods are as important as structural heterogeneity in the network when processes of disease spreading are considered \cite{de2018impact,darbon2019disease}. However, even when node-dependent recovery rates are endowed, the recovery remains a single-body type of process ($I \to S $).

In this work, we investigate the effects of dynamical recovery rates in a model of adoption dynamics on a multiplex network\cite{czaplicka2016competition}. The key feature of our model is the presence 
of a time- and node-dependent recovery mechanism that is not a spontaneous process, but depends on the states of the neighboring nodes.
Following the analogy with the processes of complex contagion, we name {\it complex recovery} the one-to-many-body recovery process of our model.
Moreover, in the model, spreading and recovery processes are implemented on the different layers of a multiplex network. Within our problem of interest, that is the adoption of a new service within the smart power grid, a model considering the local effects of neighbors seems more relevant than a simple SIS model.

In order to have a clearer perspective on smart power grids, and thus better understand the motivations behind our model, let us now spend a few words on the rapid changes that the electrical supply system is currently undergoing.
To reach the ambitious climate goals set out in Paris \cite{Paris2015}, distributed generators are being installed and centrally controlled infrastructures are being replaced by decentralized ones, so that the generation of energy can be de-carbonised.
New business models have then emerged to facilitate new modes of operation of the electricity supply, for example via concepts such as \emph{smart grids} \cite{giordano2012business,rodriguez2014business,Schaefer2015}. Within a smart grid, the different actors (agents or components) of the electrical system, ranging from fossil fuel plants and solar panels to industrial and household consumers, need to communicate and coordinate in order to allow a smooth and stable operation of the grid. 
One important instrument of a smart grid is the \emph{demand response} (DR) offered from the consumer side. Instead of consuming electricity whenever the consumer wishes, they might enter a contract, guaranteeing that a certain share of consumption will be shifted to periods of low demand. Certain consumption is easily shifted, e.g. water can  be heated and stored in hot water tanks for usage throughout the day, or electrical cars can be charged flexibly, given they are sufficiently charged for the next journey. DR can be offered in a static scheme with fixed low-demand periods, such as during the night, or it may be implemented as a dynamical scheme which constantly updates the consumption based on the actual available supply and demand by other customers.

Consumers are typically motivated to follow the DR scheme by price incentives.
Previous studies based on game-theory and optimization approaches have shown that time-varying prices might be able to align the optimal schedule of individual power consumption with the global optimum of the system \cite{li2011optimal}. If prices are also based on consumption level, these mechanisms can be efficiently used by single companies via scheduling games in order to minimize energy costs \cite{mohsenian2010autonomous}. Other studies have investigated the effects of increasing participation in DR schemes on the different market participants \cite{su2009quantifying}.

An important point is that, in order for DR schemes to be effective, a sufficiently large share of households is required. First, any business addressing households will not be profitable if only a very few participate. Even more importantly, large groups of consumers could act similarly to a virtual power plant \cite{pudjianto2007virtual} by providing power via demand control as a service to the grid. This is specifically profitable if many customers in a given region are part of the contract and can provide power within one distribution grid branch. 
Previous studies have found that consumers require positive feedback to stay within demand control contracts \cite{siano2014demand}. Hence, agents, i.e. customers opting into demand control contracts should be rewarded, e.g. by being paid a share of their contribution towards stabilizing the grid and reducing operational costs. Since  large clusters of local consumers can act easily as a virtual power plant, we assume that rewards for agents geographically surrounded by other agents opted into the contract could be higher.

In our work, we study the dynamics of signing contracts under DR schemes by modelling the system as a multiplex network, where the social layer of the customers and the layer of physical connections among households (as given by the power grid at the distribution level) are considered at the same time and coupled together. The adoption dynamics driving the contract signature is based on social influence mediated one-to-one social interactions. Therefore, we make use of epidemic spreading on the social network, where the contagion process consists in the standard simple contagion (modelling the word of mouth). Contrarily, the recovery probability depends on the local dynamics on the power-grid layer where economic incentives are implicitly included. The basic idea we want to model here, is that a power supplier will benefit from having a cluster of individuals who signed the contract within a localized geographical area, and in turn it will provide a better offer to the customers. This additional benefit, combined with the social effect of being surrounded by agents of the same type, will make the customers who signed less keen to opt out.

This paper is structured as follows: In Sec.~\ref{sec:model}, we introduce the {\em Adoption Dynamics Model} (ADM), explaining in particular the presence of a {\em complex recovery} (CR) mechanism in the model, and its motivation. In Sec.~\ref{sec:MF}, we present analytic results of the ADM in a mean-field approximation. In Sec.~\ref{sec:results_synthetic}, we discuss the results of numerical simulations of the model on two synthetic network structures, namely a duplex formed by two Erd\H{o}s-R\'enyi random graphs, as well as another duplex consisting of a small-world network and a regular 2D lattice. 
In Sec.~\ref{sec:results_real}, we focus on the application to the smart grid by using the street network as a proxy for the power grid network at the distribution level. Although not entirely representative of the real distribution of electricity, such a network encodes the geographical proximity of the households, thus it provides a more realistic representation.
Finally, we investigate the effects of the initial conditions on the temporal dynamics of the model. Conclusions and future perspectives are summarized in Sec.~\ref{sec:summary}. 

\section{The adoption dynamics model}\label{sec:model}

\begin{figure*}[t]
	\centering
	\includegraphics[width=0.8\textwidth]{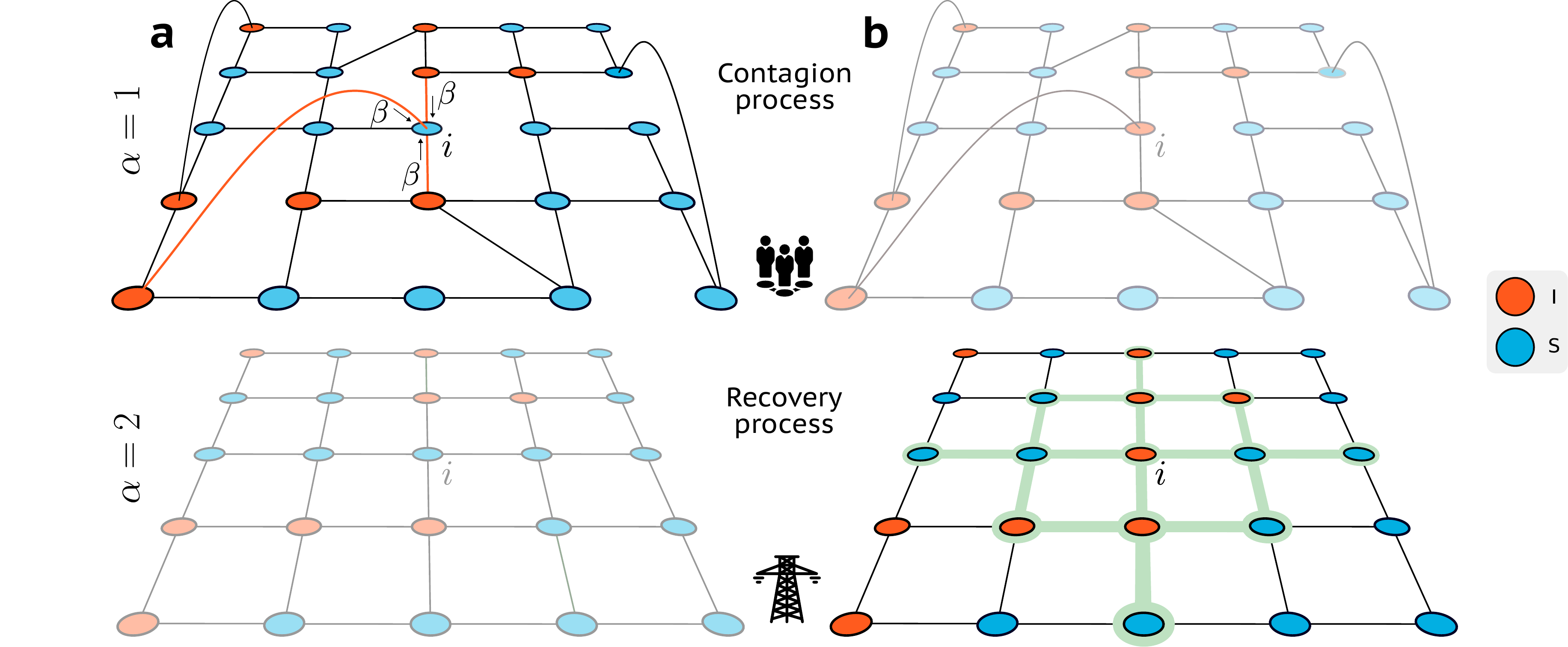}
	\caption{\label{fig:duplex} Illustration of the Adoption Dynamics Model (ADM). The two layers of the   multiplex network stand for the social ties (layer $1$, top) and connections in the power grid (layer $2$, bottom). Susceptible and infected nodes are colored in blue and red respectively. (a) The spreading dynamics takes place on the layer 1 according to a standard mechanism of simple contagion, where a susceptible node $i$ can get infected by each one of its infected neighbors with an independent probability $\beta$. (b) Contrarily, an infected node recovers with a node-dependent and dynamically changing recovery probability, which depends on the states of the neighbors at layer $2$. The shaded regions highlighted in green indicate the subset of nodes at distance $h$ hops from $i$ (the case $h=2$ is shown here), which are considered for the computation of the dynamical recovery rate $\mu_i (t)$ [see Eq.~\eqref{eq:mu_h}].}
\end{figure*}

Our model of adoption dynamics is formulated in terms of a multilayer network framework \cite{de2013mathematical, boccaletti2014structure, kivela2014multilayer, battiston2017new,bianconi2018multilayer}. In particular, we consider a multiplex network $\vec{G} = \{(\mathcal{V},\mathcal{E}_{\alpha})\}_{\alpha=1,2}$ formed by two layers (a duplex network), composed by $N=|\mathcal{V}|$ nodes and $K_{\alpha}=|\mathcal{E}_{\alpha}|$ links. Every node $i=1,...,N$ represents a household, and it has an identical replica $\left(i,\alpha\right)$ at each layer $\alpha$. Contrarily, the nodes interact in different ways, according to the specific layer, and have different structural patterns. 
In particular, the two layers represent the following two different types of 
interactions:

{\it Social layer}.---The top layer (layer $\alpha=1$) represents the social network among the individuals which are living in the households -or street areas- that we consider to be nodes. The topology of this layer is described by the binary adjacency matrix $A^{[1]}\equiv\{a^{[1]}_{ij}\}$, whose non-zero entries represent existing social links.
We denote as $k_i^{[1]}=\sum_j a^{[1]}_{ij}$ the degree of the node $i\in\mathcal{V}$ at layer $1$, so that $\langle k^{[1]} \rangle$ gives the average degree of this layer. 

{\it Power-grid layer}.---The bottom layer (layer $\alpha=2$) represents the physical connections among households as given by the power grid at the distribution level. While the nodes are the same as the nodes of layer 1, the connections are described by another binary adjacency matrix $A^{[2]}\equiv\{a^{[2]}_{ij}\}$. Analogously to the case of the first layer, 
we denote as $k_i^{[2]}=\sum_j a^{[2]}_{ij}$ and $\langle k^{[2]} \rangle$ 
the degree of node $i$ and the average degree at layer 2.

Notice that as many other infrastructural networks, the power-grid layer can be represented as a spatial network \cite{barthelemy2011spatial}, where the nodes (a household or a street in this case) and the links are embedded in a geographical space. The connectivity of the social layer is usually more complex. In fact, it is reasonable to assume that social ties are present at two different levels: the first one is related to physical proximity, which brings neighbors to interact more, while the second level includes long-range social links connecting people living in different areas of the city or in other cities.

Our purpose is to build a model of the dynamics of signing contracts under DR schemes, which we will name the {\em Adoption Dynamics Model} (ADM). 
Hence, at each time $t$, each node $i$ of the network is characterized by a binary state variable $x_i (t)\in\{0,1\}$. 
Such a variable represents the state of household $i$ with respect to the contract at 
time $t$, with 1 indicating the user has signed a contract, and 0 indicating the user has not signed a contract yet, or has opted out. Nodes change their states according to a Susceptible-Infected-Susceptible (SIS) dynamics that takes place over the links of the first layer. We assume that the  states 0 and 1 correspond respectively to the susceptible (S) and infected (I) 
states of the SIS.

In this way, each node represents a group of individuals living in a household, and each edge of the social layer stands for a social connection along which the infection can spread, i.e. a susceptible node can opt in being convinced by one of its social links.
Each susceptible node has as many channels of infection per unit time as the number of infected neighbors at the social layer $1$. The transition $S+I\xrightarrow{\beta} 2I$ is determined by the transmission rate $\beta$, which enters directly in the pairwise interactions between susceptible and infected nodes. In our model, the parameter $\beta$ can be seen as a measure of the social or advertising pressure that convinces customers to opt into a contract.
In this way, 
the probability $p_i(t)$ of a node $i$ to get infected at time $t$ reads as

\begin{equation}
p_i(t)=  1 -  \prod_{j}  \large[ 1 - \beta a^{[1]}_{ij} x_j(t) ],
\end{equation}

where the product on the right hand side gets contributions from all the infected neighbors of node $i$ at the social layer 1, and is equal to the probability that node $i$ is not infected by any of its infected neighbors. 
\\

The transition $I\xrightarrow{\gamma_i(t)} S$ is controlled by the parameter $\gamma_i(t)$, which represents the probability that node $i$ recovers at time $t$, becoming susceptible again. Instead of the spontaneous recovery, the 1-body process typically adopted in the modelling of infectious diseases, here we consider a {\em complex recovery} (CR) mechanism, which is a many-body process. 
Namely, instead of using a constant recovery probability $\mu_0$ equal for all nodes, here we introduce a time-dependent recovery probability $\gamma_i(t)$ which can also vary from node to node. In particular, we assume that $\gamma_i(t)$ is a function of the properties of the neighborhood of node $i$ at time $t$ at the power-grid layer $2$. In this way we want to model that individuals are less likely to opt out of a contract with a specific energy supplier if their neighbors, in the power grid, have signed a contract with the same company. This can be seen as an effect of a particular bonus that an energy supplier is able to offer to an individual which is part of a cluster of customers.
We thus implement the CR by defining $\gamma_i(t)$ as:
\begin{equation}\label{eq:gamma_i}
\gamma_i(t)=(1-\theta)\mu_0 + \theta\mu_i(t), 
\end{equation}
where the parameter $\theta \in \left[0,1\right]$ controls for the importance of local interactions 
in the recovery transition with respect to a standard constant recovery parameter $\mu_0$.
Notice that, for $\theta=0$ no local effects are considered for the recovery, and the model corresponds to the standard SIS model with a constant recovery probability $\mu_0$. Contrarily, if $\theta=1$, the recovery is completely determined by the dynamical term $\mu_i(t)$, which is node-dependent and that co-evolves in time together with the spreading process at layer $1$.

We consider now the case in which $\mu_i(t)=\mu_{i,h}(t)$ is a function of the network hop-distance $h$ at layer $2$. Namely, we define $\mu_{i,h}(t)$ as
\begin{equation}\label{eq:mu_h}
\mu_{i,h}(t)=\Biggl( 1-\frac{|\mathcal{I}_{i,h}^{[2]}(t)|}{|\mathcal{N}_{i,h}^{[2]}|}\Biggl)\mu_0,
\end{equation}
where $\mathcal{N}_{i,h}^{[2]}\subseteq\mathcal{V}$ is the set of nodes of $\vec{G}$ which are within $h$ hops 
from $i$ on layer $2$, and $\mathcal{I}_{i,h}^{[2]}(t)=\mathcal{N}_{i,h}^{[2]}\cap \{j\in\mathcal{V}: x_{j}(t)=1\}$ is the subset of these nodes which are infected at time $t$ (Fig.~\ref{fig:duplex}(a) bottom panel). Notice that the highest possible recovery probability in the expression above is equal to $\mu_0$, the same as the static case, but when node $i$ is completely surrounded by infectious neighbors $\mu_{i,h}(t)$ goes to zero. This becomes clear when inserting Eq.~\eqref{eq:mu_h} into Eq.~\eqref{eq:gamma_i}, leading to

\begin{equation}\label{eq:gamma_i_complete}
\gamma_i(t)=\mu_0\Biggl( 1-\theta\frac{|\mathcal{I}^{[2]}_{i,h}(t)|}{|\mathcal{N}^{[2]}_{i,h}|}\Biggl).
\end{equation}

In the simplest case, in which $h=1$, we can write Eq.~\eqref{eq:gamma_i_complete} directly in terms of the elements of the adjacency matrix $A^{[2]}$ as

\begin{equation}\label{eq:gamma_h1}
\gamma_i(t)=\mu_0\Biggl( 1-\theta\frac{\sum_{j}a^{[2]}_{ij}x_{j}(t)}{\sum_{j}a^{[2]}_{ij}}\Biggl).
\end{equation}

Finally, we denote the density of infected (adopters) individuals at time $t$ as $\rho(t)=I(t)/N=\sum_{i=1}^N x_{i}(t)/N$, which represents our macroscopic order parameter. At time $t=0$ all individuals are susceptible, with the exception of a seed $\rho_0=\rho(t=0)\ll 1$ of infected nodes (early adopters). 
\section{Mean-field analytical results}\label{sec:MF}
The density of infected individuals and the infection threshold as function of the different control 
parameters of the ADM can be 
obtained analytically in a mean-field (MF) approximation.
The MF approximation works well under
the homogeneous mixing hypothesis, assuming therefore
that the individuals with whom a susceptible individual
has contact are chosen at random from the whole population.
Furthermore, we also assume that all individuals have approximately the
same number of contacts at each time, and that all contacts transmit
the disease with the same probability.
As a consequence, instead of considering the specific topology of the two layers, we only focus
on average degree properties, so that we can write an equation for the temporal evolution of the density of infected individuals $\rho(t)$ as
\begin{equation}\label{eq:temp_evo_rho}
d_t \rho(t) = - \langle \gamma_i(t) \rangle \rho(t) + \beta{\langle k^{[1]} \rangle} \rho(t) \left[1-\rho(t)\right].
\end{equation}
With this approach we are assuming that each node of the social network has the same degree, 
equal to the average degree $\langle k ^{[1]}\rangle$
of the social network at layer $1$. $\langle \gamma_i(t) \rangle$ denotes the average recovery probability computed over all nodes.

For the particular case in which only the first neighbors are considered
($h=1$) we can derive a MF expression for $\langle \gamma_i(t) \rangle$,
by approximating Eq.~(\eqref{eq:mu_h}) as
\begin{equation}\label{eq:mu_mf_h1}
\langle \mu_{i,h=1}(t) \rangle
\approx
\left(1-\frac{{\langle k^{[2]} \rangle} \rho(t)}{{\langle k^{[2]} \rangle}}\right)\mu_0=
\left(1-\rho(t)\right)\mu_0.
\end{equation}

Notice that if a local tree-like structure is assumed for $A^{[2]}$, the same MF approximation would hold for any $h$,

\begin{equation}\label{eq:mu_mf_h}
\langle \mu_{i,h}(t)  \rangle \approx
\left(1-\frac{{\langle k^{[2]} \rangle}^{h} \rho(t)}{{\langle k^{[2]} \rangle}^{h}}\right)\mu_0=
\left(1-\rho(t)\right)\mu_0.
\end{equation}

Using these results for Eq.~(\eqref{eq:gamma_i}) and by substituting $\langle \gamma_{i,h}(t) \rangle$ into Eq.~(\eqref{eq:temp_evo_rho}) we can write the final MF expression for the temporal evolution of the density $\rho(t)$ of infected nodes, which reads as

\begin{equation}\label{eq:MF}
d_t \rho(t)= - \mu_0 [1-\theta\rho(t) ] \rho(t) + \beta{\langle k^{[1]} \rangle} \rho(t) [1-\rho(t)].
\end{equation}

After defining $\lambda=\beta {\langle k^{[1]} \rangle}/ \mu_0$ and rescaling the time as ${t'}=\mu_0(\lambda-\theta)t$, we obtain the equivalent equation: 
\begin{equation}\label{eq:MF_compact}
	d_{t'} \rho(t')= \rho(t')(\rho^{*}_2-\rho(t')),
\end{equation}

with 

\begin{equation}\label{eq:MF_rho2}
	\rho^{*}_2=\frac{\lambda -1}{\lambda - \theta}.
\end{equation}

The associated steady state equation $d_{t'} \rho(t')=0$ has therefore up to two acceptable solutions in the range $\rho\in [0,1]$: a trivial solution $\rho^{*}_1=0$, corresponding to the absorbing state in which there is no epidemic (no adopters) and all nodes have recovered; and a non-trivial solution $\rho^{*}_2$ which depends on the parameters of the model as follows:
\subsection{Case $\mathbf{\theta=0}$}
Let us first consider the case $\theta=0$ in which local dynamical effects are neglected. This case corresponds, as expected, to the standard SIS model, thus we recover the solution $\rho^{* [\theta=0]}_2$ that reads as

\begin{equation}\label{eq:MF_rho2_theta0}
	\rho^{* [\theta=0]}_2=1-\frac{1}{\lambda}=1-\frac{\mu_0}{\beta{\langle k^{[1]} \rangle}}.
\end{equation}

The solution $\rho^{* [\theta=0]}_2$ is acceptable, i.e. non-negative, when $\lambda\geq 1$, recovering in this way the standard epidemic threshold $\lambda^{[\theta=0]}_{c} = 1$.

Linear stability analysis shows that the solution $\rho^{*}_1=0$ is stable only when $\lambda<\lambda^{[\theta=0]}_{c}$. Contrarily, for values of $\lambda \geq \lambda^{[\theta=0]}_{c}$, the absorbing state $\rho^{*}_1=0$ becomes unstable while $\rho^{* [\theta=0]}_2$ becomes stable, i.e., the epidemic takes place.
\subsection{Case $\mathbf{\theta=1}$}
Let us consider now the other extreme case, $\theta=1$, in which only the local effects are considered in the CR, and therefore the recovery phase is purely dynamical. Also in this case the second solution of the stationary state equation becomes trivial, and reads $\rho^{*[\theta=1]}_2=1$. Thus, the system presents two stationary solutions, and it is easy to show that their stability changes at the same epidemic threshold as for $\theta=0$, so that $ \lambda^{[\theta=1]}_{c}=\lambda^{[\theta=0]}_{c}=\lambda_{c}$.
For $\lambda < \lambda_c$ $\rho^{*}_1$ is stable and $\rho^{*}_2$ is unstable, while for $\lambda > \lambda_c$ we have the opposite case. Therefore, contrarily from the completely non-local case ($\theta=0$), here the system undergoes an explosive transition from the healthy to the endemic state, where all individuals are adopters.
\subsection{General case}
In the most general case, the second solution $\rho^*_2$, given by Eq.~\eqref{eq:MF_rho2}, depends on
both the rescaled infectivity $\lambda$ (therefore on the average degree $\langle k^{[1]} \rangle$) and
on $\theta$, the parameter which controls for the local effects of the CR.
Notice that the solution is acceptable if $\rho^*_2 \in [0,1]$, which implies again $\lambda\in [1,\infty]$. Therefore, for any $\theta$, the transition from the healthy to endemic state happens at the same epidemic threshold $\lambda_c=1$, but the density of infected in the endemic state varies with $\theta$. The stability of the fixed point $\rho^{*}_2$ can be easily investigated by defining the second term in Equation \eqref{eq:MF} as $F(\rho)=\lambda \rho [1-\rho]$ and then checking the sign of the  derivative of $F(\rho)$. Since  $F'(\rho)|_{\rho=\rho^{*}_2}=1-\lambda$ does not depend on $\theta$, $\rho^{*}_2$ always represents a stable fixed point for the dynamics.
\section{Numerical results on synthetic networks}\label{sec:results_synthetic}
We present here numerical simulations of the ADM on various synthetic duplex networks. In each case   
the simulations are performed for different realizations of the networks, stopping each run whenever an absorbing state is reached. Alternatively, if a stationary state is reached the stationary density of infected is computed by considering an average over the last 100 time-steps. Each run starts with different initial conditions, given by randomly placing a seed of $\rho_0$ infectious nodes (usually $1\%$ of the nodes), and then we average the results over all the runs. Throughout all the numerical simulations presented in this paper, we restrict for simplicity to the case $h=1$. 

\begin{figure}
	\centering
	\includegraphics[width=0.48\textwidth]{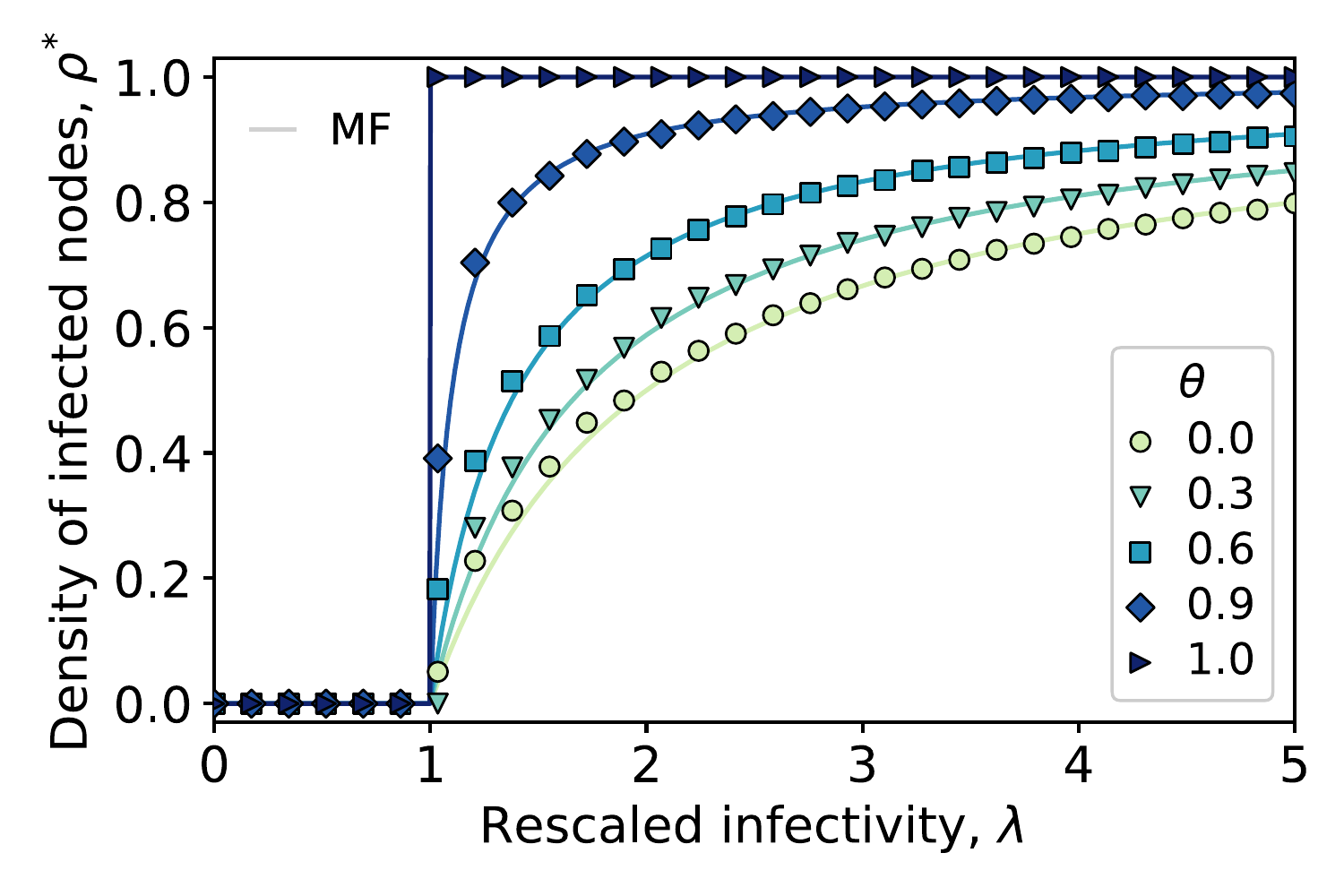}
	\caption{\label{fig:CR_on_ER} Numerical simulations of ADM ($h=1$) on a duplex network formed by two ER networks with $N=900$ and $\langle k \rangle =10$. The average fraction of infected nodes is plotted against the rescaled infectivity $\lambda=\beta\langle k^{[1]} \rangle / \mu_0$. Different curves (and colors) correspond to different values of the parameter $\theta$, which controls for the strength of the local effects in the complex recovery process, as defined in Eq~\eqref{eq:gamma_i}. The case $\theta = 0$ corresponds to the standard SIS model. Simulations (points) are plotted together with the analytical mean-field (MF) solution of Eq.~\eqref{eq:MF} (continuous lines).}
\end{figure}

The first system we have considered is a duplex with $N=900$ nodes formed by two Erd\H{o}s-R\'enyi (ER) random graphs having average degrees $\langle k ^{[1]}\rangle=\langle k ^{[2]}\rangle=\langle k \rangle=10$. Fig.~\ref{fig:CR_on_ER} shows the stationary density of infected $\rho^*$, obtained by averaging the prevalence curves for different realizations of the numerical simulations,  
as a function of the rescaled infectivity $\lambda=\beta\langle k^{[1]}\rangle/\mu_0$. 
Different curves correspond to different values of the parameter $\theta$, which controls for the local effects in the CR process. Indeed, the case $\theta=0$ is equivalent to the standard SIS model with spontaneous recovery, where a non-zero density of infected nodes in the stationary state appears for values of $\lambda$ larger than a critical value $\lambda_c=1$. 
By increasing $\theta$, the density of infected nodes in the endemic state $\rho^{*}>0$ increases and the transition becomes steeper and steeper, until the extreme case $\theta=1$. In this latter case, i.e., when the recovery is purely dynamical, the systems undergoes a discontinuous transition from the absorbing state $\rho^{*}=0$ with no adopters to the opposite state $\rho^{*}=1$. Notice that the transition occurs at the same critical threshold $\lambda_c=1$. Fig.~\ref{fig:CR_on_ER} also shows the continuous curves representing the analytical prediction in the MF approach, as given by Eq.~\eqref{eq:MF_rho2}. The match between curves and points confirms the accuracy of the MF approximation in reproducing the dynamics of the ADM in the case of random graphs, and also its ability to capture the different types of transitions the ADM exhibits when 
the value of $\theta$ is changed.

\begin{figure}
	\centering
	\includegraphics[width=0.48\textwidth]{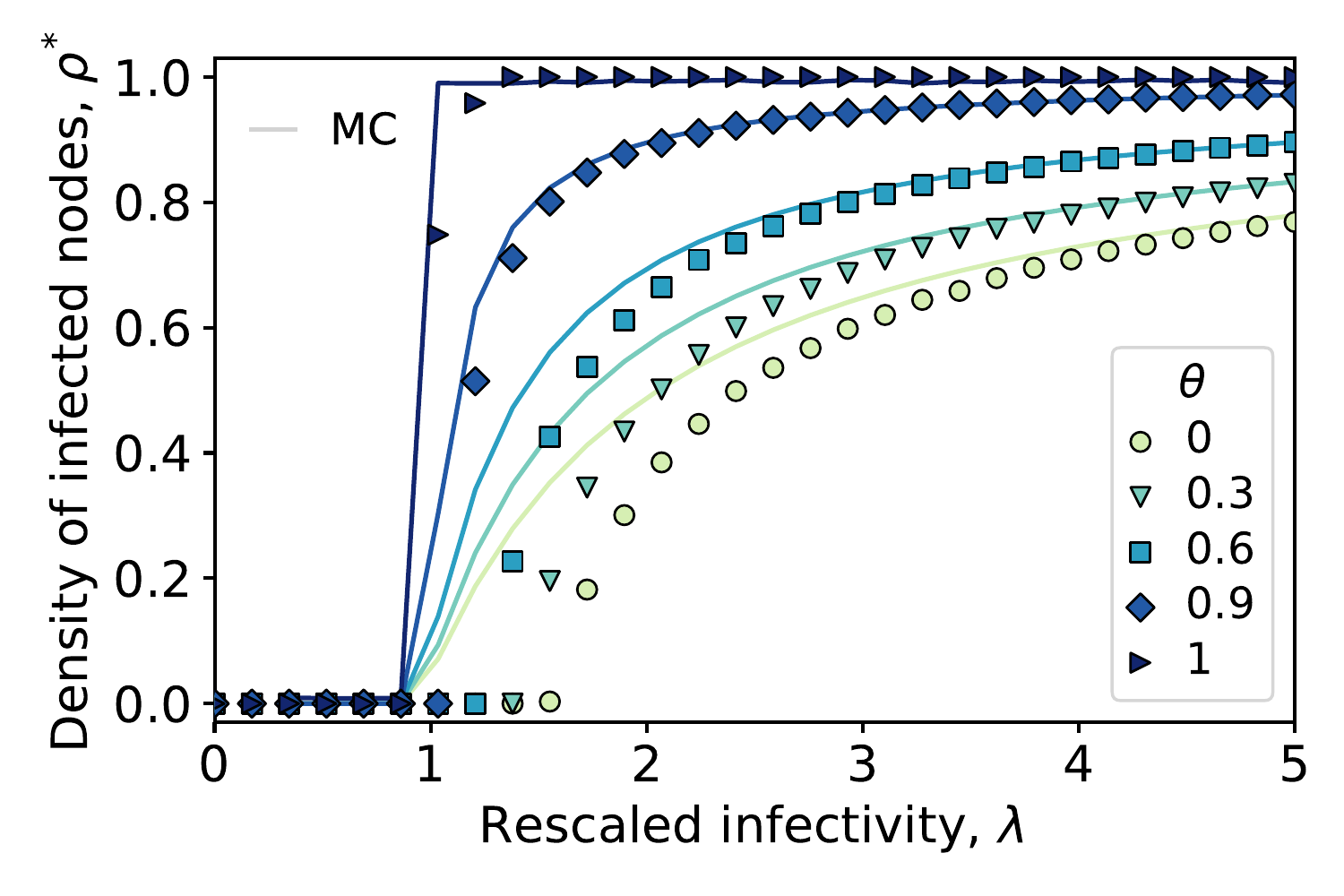}
	\caption{\label{fig:CR_on_SW} Numerical simulations of the ADM ($h=1$) on a duplex network formed by a 2D lattice and a SW network with $N=2500$ nodes, $\langle k^{[2]} \rangle\approx 4$, and $p=0.01$. The average fraction of infected nodes is plotted against the rescaled infectivity $\lambda=\beta\langle k^{[1]} \rangle / \mu_0$. Different curves (and colors) correspond to different values of the parameter $\theta$. Simulations (points) are plotted together with the curves obtained with the discrete-time Markov chain (MC) approach as given by Eq.~\eqref{eq:rho_Markov} (continuous lines).}
\end{figure}

As a second system, we have considered a slightly more realistic synthetic duplex network. In particular, we model the power-grid layer as  a 2D lattice ($N=2500$, $\langle k^{[2]} \rangle\approx 4$) and we couple it to a social layer which is obtained 
from the same 2D lattice, by 
rewiring each of its links at random with a probability $p=0.01$. It is worth clarifying that we will call this layer small-world (SW), given the similarity of the rewiring mechanism with the original small-world model proposed by Watts and Strogatz \cite{watts1998collective}. The rewiring mechanism, adopted only at layer 1, breaks the regularity of the lattice by introducing social connections between nodes that are not first neighbors at the level of the power grid network in layer 2. The results obtained are shown in Fig.~\ref{fig:CR_on_SW}. We notice a few differences with respect to the results reported in Fig.~\ref{fig:CR_on_ER}. In particular, we observe that the threshold $\lambda_c$ slightly increases when the value of $\theta$ changes from $\theta=0$ to $\theta=1$, and this behaviour is not captured neither by the analytical predictions in the MF approximation, nor by a more accurate discrete-time Markov chain (MC) approach (see Appendix \ref{appendix:markov}), whose curves are shown as continuous lines. Such differences might be due to the strong correlations between nodes induced by the underlying lattice-like structure of SW networks and to the limitations of the MC approach caused by the time discretisation~\cite{fennell2016limitations,valdano2018epidemic}. 
\section{Numerical results on real-world networks}\label{sec:results_real}

\begin{figure*}[t]
	\centering
	\includegraphics[width=0.95\textwidth]{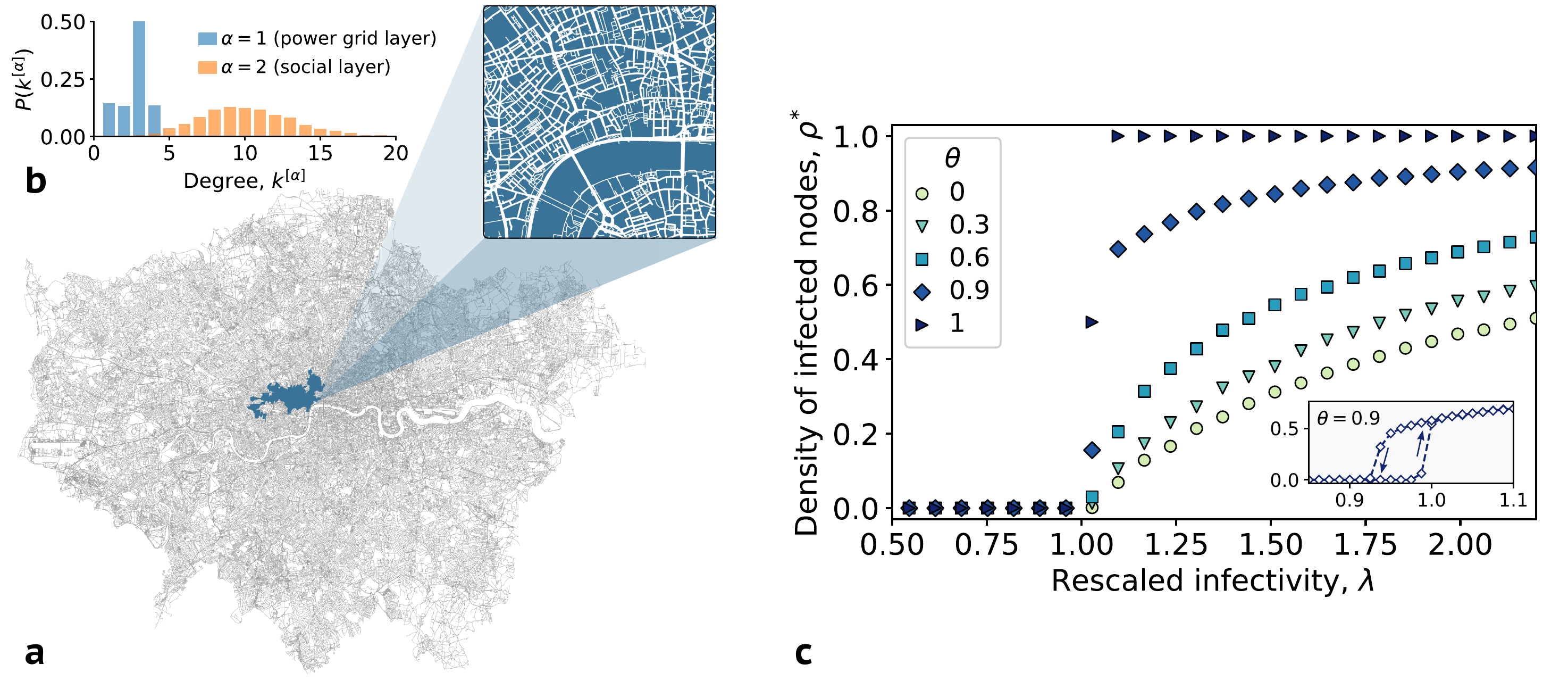}
	\caption{\label{fig:London} ADM ($h=1$) on a real-world duplex network in which a street network is used as a proxy for the power grid. (a) A central neighborhood in London is selected by using a hierarchical percolation approach (blue zone). The degree distributions of the street network and the coupled social network constructed from it are shown in panel (b). (c) The average fraction of infected nodes obtained by means of numerical simulations is plotted against the rescaled infectivity $\lambda=\beta\langle k^{[1]} \rangle / \mu_0$. Different curves (and colors) correspond to different values of the parameter $\theta$, which controls for strength of the local effects in the CR process, as defined by Eq~\eqref{eq:gamma_i}. The inset shows the hysteresis loop, which  appears close to the threshold for $\theta=0.9$.}
\end{figure*}

In the previous section we explored the model on two synthetic duplex networks. With the first we observed the phenomenology on two random graphs, while in the second we considered a more realistic -yet synthetic- structure composed by a lattice and a SW network. Here, we make a further step in this direction by considering street networks \cite{porta2006network} from the real world as proxies for power grid networks at the distribution level, and a multilayer adaptation of the well-known Waxman random graph model\cite{waxman1988routing} to represent the social layer. With this approach, the spatial nature of the street network is used both to embed the power grid into the physical space and to shape the connectivity patterns of the social layer. Due to the use of the Waxman model, the social connections decay exponentially with the distances on the network, which in turn are affected by the physical constraints imposed by the morphology of the territory. More details on the procedure to construct the street network layer and the social layer are given in Appendix~\ref{appendix:real_mux}.

We restrict our attention to a neighborhood of the city of London [see Fig.~\ref{fig:London}()a)], defined by using a hierarchical percolation approach, as introduced in Arcaute et al.~\cite{arcaute2016cities}. The method gives rise to neighborhoods at different scales arising from the density of the street intersections. In the case of London, some scales reveal its composition in terms of historical villages, corresponding now to differentiated neighborhoods. By selecting an appropriate scale, this method allows us to focus on a relatively small portion of the city, as a targeted adoption campaign would do, while keeping at the same time the computational cost at a reasonable level. For the details of the method see Appendix~\ref{appendix:real_mux} and Ref.~\cite{arcaute2016cities}. The resulting duplex network has $N\approx 3000$ nodes, with an average degree of $\langle k^{[1]}\rangle\approx 10$ at the social layer and $\langle k^{[2]}\rangle\approx 3$ at the power-grid layer. The associated degree distributions are shown in Fig.~\ref{fig:London}(b).
As in the previous cases, we investigate the density of infected individuals in the stationary state as a function of the rescaled infectivity. The plots reported in Fig.~\ref{fig:London}(c) confirm similar results to those obtained with synthetic networks. In particular, a clear change in the nature of the transition is observed also when more realistic network structures are used both at the grid and at the social layer.  Associated to the sudden transitions at large values of $\theta$, we have also observed the appearance of hysteresis loops. An example is shown 
in the inset plot for the case  $\theta=0.9$. 
In the next section we will explore these phenomena more in details.

In what follows we briefly investigate the effects of initial conditions in the evolution of the density of infected nodes \cite{karampourniotis2015impact}. Most of the existing literature targets this problem within the domain of infectious diseases spreading, which translates into looking for optimal immunisation strategies, i.e., key nodes to vaccinate in order to limit the spread \cite{pastor2002immunization, kitsak2010identification, morone2015influence}. Here, we investigate the temporal aspect of the infection dynamics as a function of the initial conditions. This will be done in two different ways, since we can control for both the number and the position in the network of the initial adopters, i.e. of those nodes who will initiate the spreading. 

\begin{figure*}[t]
	\centering
	\includegraphics[width=1\textwidth]{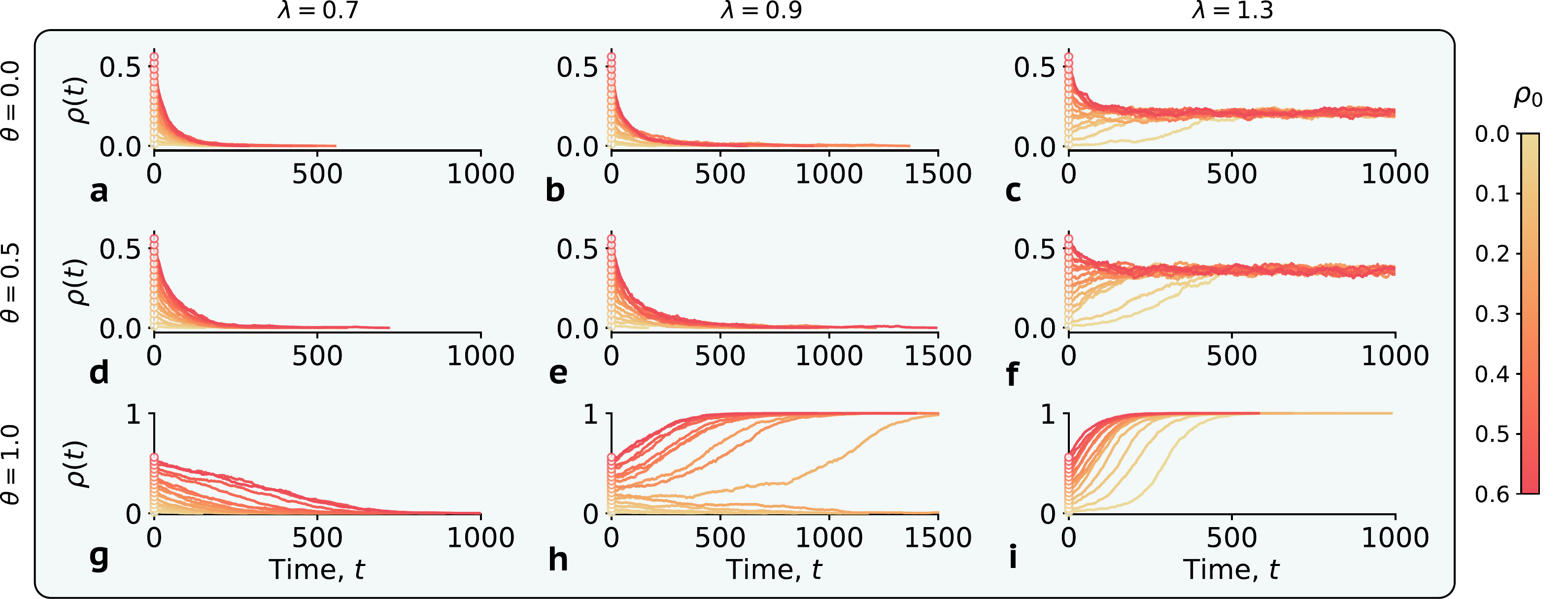}
	\caption{\label{fig:London_initial_conditions} Effect of the initial density of adopters on the temporal evolution of the spreading. Each panel shows  
		the densities of infectious nodes for different sizes of the initial seed of infectious adopters $\rho(t=0)=\rho_0$ and for different values of ($\theta, \lambda$). (a)-(c) refer to the standard SIS model, without local effects, while (g)-(i) represent the other extreme case in which the recovery process is completely controlled by the local dynamics. Different scales have been adopted for panels (b), (e), and (h) due to the proximity of the epidemic threshold $\lambda_c$, which makes the runs last longer.}
\end{figure*}

\subsection{Varying the size of the initial seed}

\begin{figure*}[t]
	\centering
	\includegraphics[width=0.9\textwidth]{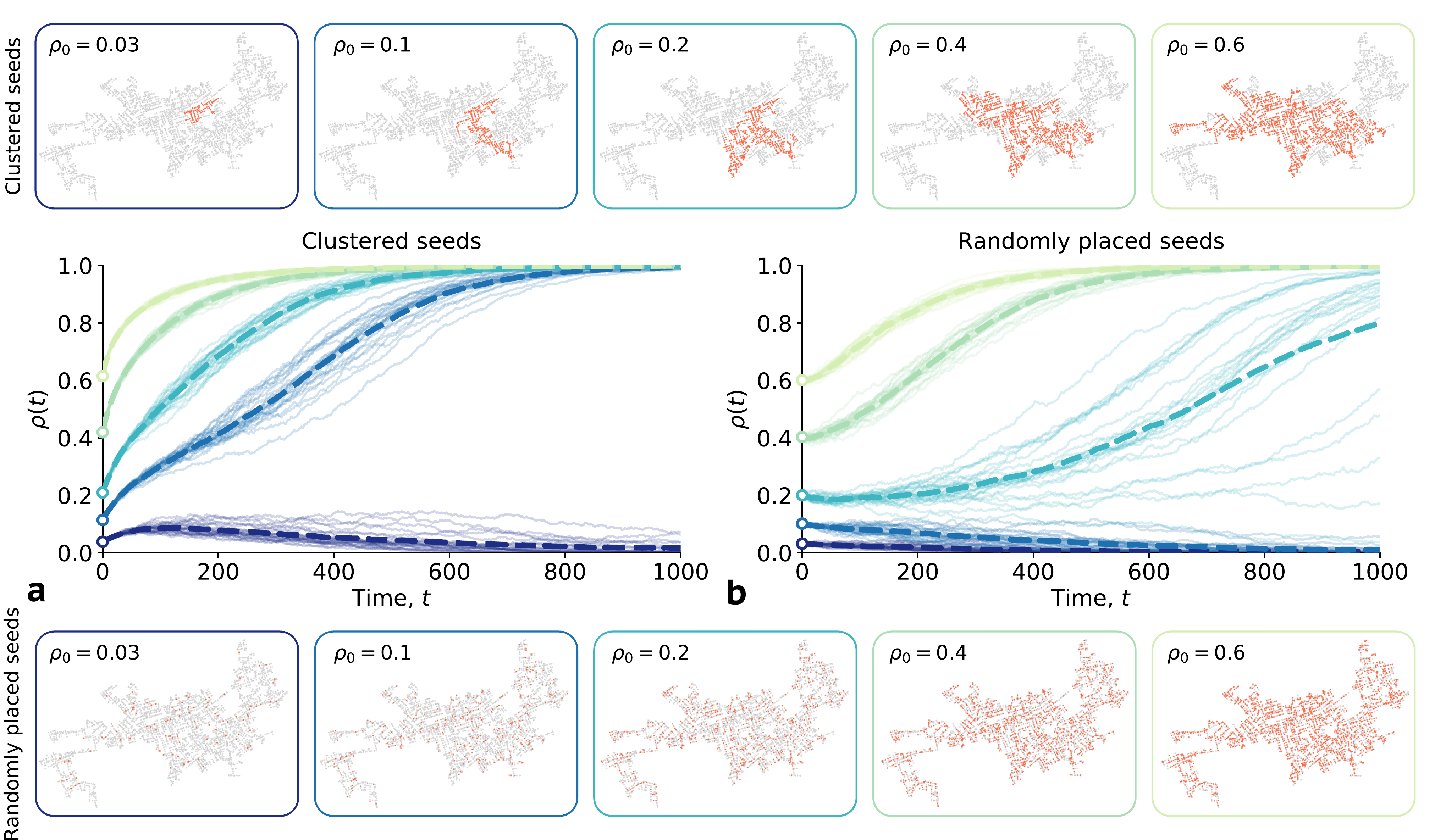}
	\caption{\label{fig:London_initial_placement} Effect of the position of the initial seed of adopters in the ADM on the real-world duplex network with parameters: $\lambda=0.9, \theta=1$. The temporal evolution of the densities of infectious nodes is showed for the two considered scenario: (a) a clustered seed of infectious nodes on the power-grid layer and (b) a randomly placed seed of infectious nodes. Different colors correspond to different sizes of the initial seed of infectious $\rho_0$ (single realizations are plotted as continuous lines, while dashed lines represent their average). The actual positions of the seeds are shown, for each $\rho_0$, in the top and bottom maps, representing, respectively, the clustered and the random scenario. Infectious nodes are depicted in red.}
\end{figure*}

We start considering a set of randomly placed infectious nodes, as before, whose size at time 0 is controlled by the density $\rho(t=0)=\rho_0$. We then simulate the ADM with different values of the parameters $(\lambda, \theta)$ for different initial densities $\rho_0$ in the range $(0,0.6]$. Results are shown in Fig.~\ref{fig:London_initial_conditions}. Each panel corresponds to a given pair of parameters $(\lambda, \theta)$, while different curves within the same panel display the temporal evolution of the density of infected nodes when considering different $\rho_0$ (see colorbar on the right-hand-side). Rows indicate different values of $\theta$, moving from the standard SIS model with no dynamical recovery ($\theta=0$, a-c), to the other extreme case in which the recovery process completely depends on the local dynamics of the neighboring nodes ($\theta=1$, g-i). An intermediate case with $\theta=0.5$ is also considered (panels d-f). Similarly, we use three values of the infectivity $\lambda$: one below the epidemic threshold ($\lambda=0.7$, a,d,g), one close to the epidemic threshold ($\lambda=0.9$, b,e,h), and one above the epidemic threshold ($\lambda=1.3$, c,f,i). Trivial effects are found when we are below and above the threshold. In particular, in the first case (panels a,d,g), the higher the $\rho_0$ the longer it takes to the system to reach the absorbing state $\rho^*=0$. Similarly, when high values of $\lambda$ are considered (panels c,f,i) what matters is the distance between the initial density of the seed $\rho_0$ and the final stationary state, which in turn depends on $\theta$. However, close to the threshold, the strong dynamical effects of the CR process create a bi-stable region (panel h), in which the initial density of infectious nodes $\rho_0$ determines whether the systems will end up in the absorbing states without adopters ($\rho^*=0$) or with all adopters ($\rho^*=1$). Notice that such bi-stability is not present in the MF formulation presented in Sec.~\ref{sec:MF}.

\subsection{Varying the position of the initial seed}

To better understand the phenomenology of the ADM  close to the threshold, we fix the parameters to the latest case ($\lambda=0.9$, $\theta=1$) and explore the effects of the initial position of the seed, while still varying its size. To do this, we consider a different scenario in which the initial adopters form a cluster in the power-grid layer. This cluster corresponds to a smaller neighborhood within the considered area, and it is selected with the same hierarchical percolation approach used in the construction of the real-world multiplex network, but for smaller percolation thresholds (see Appendix \ref{appendix:real_mux} for more details). We will compare this scenario with 
the standard case in which the initial seed of adopters is placed at random instead.

The results are shown in panel a and b of 
Fig.~\ref{fig:London_initial_placement} respectively. In each case the temporal evolution for the density of infected nodes is plotted for different values of $\rho_0$ (different colors). Shaded curves are different realizations of the ADM, while their average is represented by the dashed thick lines. As a reference, we also add the maps showing the exact position of the seed of infectious nodes (depicted in red) for each scenario and for the different values of $\rho_0$ considered. Maps for the clustered seed and randomly placed seed are shown in the top and bottom row respectively. Different behaviours emerge from the two scenarios. Indeed, the prevalence curves seem to fluctuate more when the seed is placed at random (panel b).
Interestingly, this phenomenon is in contrast with the results of the MF formulation and the simulations of the ADM on synthetic ER networks discussed in Sections~\ref{sec:MF} and~\ref{sec:results_synthetic}.
As a result, when a density $\rho_0=0.2$ of initial adopters is considered, fluctuations are so strong that, even with exactly the same (random) initial positions of adopters, the system can end up in either the absorbing state  $\rho^*=0$ or in the state $\rho^*=1$. If clustered seeds are considered instead (panel a), the same seed size $\rho_0=0.2$ always drives the system to the absorbing state with all adopters ($\rho^*=1$). In this scenario, the critical value for $\rho_0$ which separates the two basins of attractions seems to be better defined: see the curves for $\rho_0=0.03$ and $\rho_0=0.1$. Finally, it is worth noticing that for this last seed size, $\rho_0=0.1$, the initial placement completely determines the final state of the system. Indeed, only $10\%$ of clustered adopters are enough to drive the system towards the full adoption case, while this does not happen if a seed of the same size is placed at random.

\section{Summary and conclusions}\label{sec:summary}

In this work we have introduced and studied both numerically and when possible 
analytically a mathematical model  
of spreading on a network with a dynamical recovery mechanism of the nodes, which is a function of the network state. Our original purpose is to reproduce the dynamics of service adoption in demand response management~\cite{bale2013harnessing,rai2015agent,rai2016agent,hesselink2019adoption}, in which the behaviour of a customer is influenced by its social contacts, in addition of also depending on the specific spatial configuration of other customers in close proximity within the power grid service area. For this 
reason, we consider a duplex network with a social layer and a power-grid layer.  The adoption process is modelled as an epidemic spreading on the social layer, with a recovery rate of the nodes that depends on the states of their neighbors on the power-grid layer. In this way the dynamics tends to preserve clusters of infected individuals by making an infected node surrounded by nodes in the same state less keen to recover. 

 Results suggest that the more the recovery of the nodes depends on the local influence of peers (large values of $\theta$), the more discontinuous the transition from non-adopters (healthy) to full adoption (fully infected network) becomes. 
In simulations on real-world networks, such as the London network, we also noted that the final state of the system is not uniquely defined by the infection and recovery parameters $\beta$ and $\mu$, but the initial conditions can have a substantial impact on the spreading dynamics, with clustered seeds in the power-grid layer more likely leading to full network infection. 
We have found that a mean-field approximation captures the simulation results nicely for Erd\H{o}s-R\'enyi networks, while more advanced analytical descriptions might be necessary to characterise our model on more realistic and complex network structures.

While we motivated our model from the electricity demand management, other applications that rely on local customer resources should follow similar dynamics\cite{nyborg2016social}. This could for example include car sharing or citizen science projects. The main message we can derive for all such systems from the analysis of our model is the following. In any real application case, we would first need to determine the strength of the local influence of other customers, i.e. the magnitude of the parameter $\theta$ in our model. If such local influence is weak, a smooth transition to a non-zero 
density of adopters takes place when the infectivity is above a given threshold. When the local influence is strong instead, we observe a sudden transition and the appearance of an intermediate (hysteresis) region where, as soon as a critical mass is reached, (almost) everyone would adopt the new technology. 

Our results also show that strong local influence is key to determine adoption, hence giving insights on how to strategically plan on the nodes to be targeted initially. Namely, reaching out to customers who are physically spread out in the power-grid layer, or explicitly targeting social clusters which are not defined spatially, is more likely to fail than starting the adoption process from clusters in the power-grid layer, which provides the positive feedback-loop for the consumers. In fact, a cluster of "adopters" on the power-grid layer is likely to stay within the contract and will also convince their neighbours to join. Therefore, this strategy will in most cases lead to a higher penetration of the new technology. Advertisement should take this into account, e.g. by explicitly advertising within local communities.
Alternatively, businesses might try to re-shape the infection and recovery process itself. Already commonly adopted "hire a friend" schemes try to build a positive feedback among customers, which can directly strengthen the infection process. Our results suggest that alternative "hire a neighbor" schemes, specifically designed to target neighbors in the power-grid layer, could also positively contribute, this time by altering the recovery dynamics. 
%

The presented approach can easily be extended into multiple directions. 
On the one hand, one could further improve the accuracy of the network topologies, for example by modelling the social network as a crossover between scale-free and spatial networks, as proposed in Ref.~\cite{barthelemy2003crossover}, or by directly using real-world friendship data. Further analytical development might also be required in this direction.
On the other hand, different variations of the model are also possible. For instance, the dynamical recovery term $\mu_i(t)$ could include influences from nodes $h=2$, or more, hops away. Alternatively, it could rely on Euclidean distances calculated on the physical space, or other distance functions more realistically mapping consumer decisions. Another natural extension might involve the introduction of other, both  exogenous and endogenous,  effects as additional contributions to the adoption dynamics. These could include, for example, mass media exposure  \cite{toole2012modeling} and the explicit introduction of economic factors such as price of energy or price incentives.
Finally, in this manuscript we focused on demand control for households, specifically neglecting industrial consumers, as they follow very different rules and require independent business models which are also more likely follow different adoption schemes. Future work should  extend the present framework to non-household customers.
\begin{acknowledgments}
	I. I., B. S., C. B. and V. L. acknowledge support from EPSRC Grant No. EP/N013492/1. I. I. and E. A. acknowledge support from EPSRC Grant No. EP/M023583/1. I. I. also acknowledges support from The Alan Turing Institute under the EPSRC Grant No. EP/N510129/1.
	B. S. acknowledges support from the Cluster of Excellence Center for Advancing Electronics at TU Dresden, the German Federal Ministry for Research and Education (BMBF grants no. 03SF0472F and 03EK3055F). This project has received funding from the European Union’s Horizon 2020 research and innovation programme under the Marie Sklodowska-Curie grant agreement No 840825. V. L. acknowledges support from the Leverhulme Trust Research Fellowship 278 “CREATE: the network components of creativity and success”.
	We thank Alex Arenas, Jesus G{\`o}mez-Garde{\~n}es, Sandro Meloni
	and Benjamin Steinegger for useful comments and suggestions.
\end{acknowledgments}
%
%
\appendix
\section{CONSTRUCTION OF MULTIPLEX NETWORKS}\label{appendix:real_mux}
Here we describe the methodology used to construct the real-world inspired multiplex networks. 
We constructed the power-grid layer at the distribution level by taking as a proxy the street network in its primal approach \cite{porta2006network}, i.e., by representing crossroads as nodes connected by streets. According to our ADM, each node corresponds to a household, hence we approximate the households connectivity in the distribution grid by the street network formed by intersection points linked by streets.

The street network was constructed starting from the same data set used in Ref.~\cite{arcaute2016cities}: the Ordnance Survey (OS) MasterMap \cite{dataset}. The data set consists of a clean street network for the entire Britain, in which roundabouts have been replaced by single intersections, and in which each edge comes with an associated weight representing the length of the street (for more details see Ref.~\cite{arcaute2016cities}). We first restricted our data set to the Greater London Authority by retaining only those points (nodes) falling within the boundaries of the LSOA (Lower Super Output Area)\cite{LSOA}. Then, we selected a smaller neighborhood in central London by following the hierarchical percolation method proposed in Ref.~\cite{arcaute2016cities}. The method produces a clustering on the nodes based on a single parameter $\epsilon$, which acts as a percolation threshold on the street distance between the points. More precisely, given a threshold $\epsilon$, the graph is divided into different connected components corresponding to the sub-graphs induced by the thresholding on the nodes at a distance smaller than $\epsilon$ (see Ref.~\cite{arcaute2016cities} for more details). The resulting network, corresponding to the largest connected component obtained for a threshold of $\epsilon=89$ meters, is composed of $N=3379$ nodes and $K^{[2]}=4602$ links.

We constructed the geographical social network starting from the well-known Waxman random graph model\cite{waxman1988routing}. In the standard model, nodes are initially placed at random over a plane and then connected in pairs with a probability that decays exponentially with their distance. Here, we modified the model in two ways: (i) nodes are not placed at random, but the geographical position of each node on the social layer corresponds to the position on the power-grid layer; (ii) instead of considering the geographical distance between the nodes, we considered the network distance. Notice that, being the network embedded in space, the network distance is already shaped by the particular spatial displacement of the nodes. More precisely, the model works as follows.

Given the set of nodes (and their position), let us call $d^{[2]}(i,j)$ their distance in layer 2 (the power-grid layer). Then, the probability that $i$ and $j$ are connected in layer 1 is given by

\begin{equation}\label{eq:social_pij}
    P^{[1]}(i,j)=\alpha \exp\left[-d^{[2]}(i,j)/\alpha L^{[2]}\right]
\end{equation}

where $L^{[2]}$ denotes the diameter of layer 2 and $\alpha$ is a tunable model parameter.

We constructed the network for the social layer by linking the nodes of the grid layer according to the probability given by Eq.~\eqref{eq:social_pij}, with $\alpha$ tuned in order to obtained a reasonable number of influential household connections ($\alpha=0.003$).
The resulting network has $K^{[1]}=17183$ links, corresponding to an average of $\langle k^{[1]}\rangle \approx 10$ connections per household.

\section{DISCRETE-TIME MARKOV CHAIN APPROACH}\label{appendix:markov}

 In the discrete-time Markov chain approach, the probability of node $i$ to be infected at time $t$ $\text{Prob}\left[x_{i}(t)=1\right]=\pi_i(t)$ is a random variable, and it is assumed that for different nodes these probabilities are independent \cite{gomez2010discrete}. 
The equation for the discrete-time evolution of $\pi_i(t)$ can then be written as
\begin{equation}
\pi_i(t+1)= (1-q_i(t))(1-\pi_i(t)) + (1-\gamma_i(t))\pi_i(t) 
\end{equation}
where the two terms of the right-hand side are, respectively,

\begin{itemize}
	\item the probability that node $i$, susceptible at time $t$, gets infected by a neighbor and
	\item the probability that node $i$, infected at time $t$, does not recover.
\end{itemize}

$q_i(t)$ represents the probability of node $i$ not being infected by any of his neighbors at time $t$, and it can be written in terms of the adjacency matrix $A^{[1]}$ of layer $1$, which controls the contacts between nodes $i$ and $j$, as

\begin{equation}\label{eq:MC_qi}
q_i(t)=\prod_{j} 1-\beta a^{[1]}_{ij} \pi_j(t)
\end{equation}

while the recovery probability $\gamma_i(t)$, for the case $h=1$, is given by

\begin{equation}
\gamma_{i,h=1}(t) = \mu_0\Biggl(1-\theta\frac{\sum_{j}a^{[2]}_{ij}\pi_{j}(t)}{\sum_{j}a^{[2]}_{ij}}\Biggl) 
\end{equation}

Finally, the stationary state $\pi_i(t+1)=\pi_i(t)$ is given by
\begin{equation}\label{eq:MC_stationary}
\pi_i=(1-q_i)+(q_i-\gamma_i)\pi_i 
\end{equation}
The system of equations given by Eq.~\eqref{eq:MC_stationary} is then solved numerically, and the density of infected is obtained by taking the average over all the nodes,
\begin{equation}\label{eq:rho_Markov}
    \rho = \frac{1}{N}\sum_{i} \pi_i
\end{equation}
The limitations of this approach have been discussed in Ref.~\cite{fennell2016limitations,valdano2018epidemic}. Moreover, in our specific case, the underlying lattice-like structure of the network topologies used in Section~\ref{sec:results_synthetic} can break the assumption that the probabilities $\pi_i(t)$ and $\pi_j(t)$ of two different nodes to be infected are independent.
\section*{REFERENCES}

\clearpage

\end{document}